\documentclass[useAMS]{mn2e}
\usepackage{amsmath}
\usepackage{times}
\usepackage{graphicx}
\usepackage{aas_symbols}
\usepackage{graphicx}
\usepackage{epstopdf}

\newcommand{\beq}{\begin{equation}}
\newcommand{\eeq}{\end{equation}}
\newcommand{\bea}{\begin{eqnarray}}
\newcommand{\eea}{\end{eqnarray}}

\title[Outflow model for M101 ULX-1]{The Nature of ULX Source M101 X-1:  Optically Thick Outflow from A Stellar Mass Black Hole}

\author[R.-F. Shen et al.]{Rong-Feng Shen,$^{1,2}$\thanks{Email:rf.shen@mail.huji.ac.il} Rodolfo Barniol Duran,$^{1}$ Ehud Nakar$^{2}$ and Tsvi Piran$^{1}$\\
$^{1}$Racah Institute of Physics, Hebrew University of Jerusalem, Jerusalem 91904, Israel\\
$^{2}$Department of Astrophysics, School of Physics and Astronomy, Tel Aviv University, Tel Aviv 69978, Israel}

\begin{document}

\date{Accepted 2014 November 13.  Received 2014 November 13; in original form 2014 November 1}

\maketitle

\begin{abstract}
The nature of ultra-luminous X-ray sources (ULXs) has long been plagued by an ambiguity about whether the central compact objects are intermediate-mass (IMBH, $\gtrsim 10^3 M_{\odot}$) or stellar-mass (a few tens $M_{\odot}$) black holes (BHs). The high luminosity ($\simeq 10^{39}$ erg s$^{-1}$) and super-soft spectrum ($T \simeq 0.1$ keV) during the high state of the ULX source X-1 in the galaxy M101 suggest a large emission radius ($\gtrsim 10^9$ cm), consistent with being an IMBH accreting at a sub-Eddington rate. However, recent kinematic measurement of the binary orbit of this source and identification of the secondary as a Wolf-Rayet star suggest a stellar-mass BH primary with a super-Eddington accretion. If that is the case, a hot, optically thick outflow from the BH can account for the large emission radius and the soft spectrum. By considering the interplay of photons' absorption and scattering opacities, we determine the radius and mass density of the emission region of the outflow and constrain the outflow mass loss rate. The analysis presented here can be potentially applied to other ULXs with thermally dominated spectra, and to other super-Eddington accreting sources.
\end{abstract}



\section{Introduction}

The recurrent ultra-luminous X-ray (ULX) source X-1 in the nearby galaxy M101 (hereafter M101 ULX-1) is characterized by two states (Kong, Di Stefano \& Yuan 2004; Mukai et al. 2005; Liu 2009): the ``high states'' are transient outbursts with peak luminosities $L_X \sim 10^{39-40}$ erg s$^{-1}$, which are separated by long (typically 150 - 200 days) ``low states'' with $L_X \sim 10^{37}$ erg s$^{-1}$. High states appear to have durations of 10 - 20 days, but the amount of time spent at the luminosity peak is significantly shorter (e.g., the time spent at the peak is $\sim 2$ days for the 2004 December / 2005 January outburst). The high-state X-ray spectra are very soft and can be fitted with a blackbody with a temperature $ \simeq 0.1$ keV. The low-state spectra are relatively hard; a combination of a power law and a blackbody is needed, though the color temperature is not different from the high states (Kong et al. 2004).    

The high luminosity and low color temperature of M101 ULX-1 during the high states immediately suggest that the emission radius is at least $\sim 10^9$ cm. Therefore, this source has been considered as an intermediate mass black hole (IMBH) candidate. Recently using optical spectroscopy, Liu et al. (2013) identified the secondary in the binary system of M101 ULX-1 as a $M_*$= 19 $M_{\odot}$ Wolf-Rayet (W-R) star (subtype WN8) of radius 11 $R_{\odot}$ with a strong mass loss $\dot{M}_* \simeq 2\times10^{-5} M_{\odot}$ yr$^{-1}$. Fitting of nine radial velocity measurements gives an orbital period $P=$ 8.2 days and a low eccentricity (Liu et al. 2013). The mass function is $M^3 \sin^3 i /(M+M_*)^2= 0.18~M_{\odot}$ where $M$ is the mass of the primary (BH) and $i$ is the inclination angle. These values suggest a 20 -- 30 $M_{\odot}$ black hole (BH) as the primary orbiting at a $\sim 60 R_{\odot}$ separation from the secondary. An IMBH of $10^3~ M_{\odot}$ (300 $M_{\odot}$) would require $i=3^{\circ}$ ($i=5^{\circ}$). The probability of observing a pole-on binary with $i<3^{\circ}$ ($i=5^{\circ}$) is less than 0.001 (0.003). 

However, the large emission radius $\gtrsim 10^9$ cm poses a problem to any model of an accretion disk around a stellar mass BH: It is much larger than the disk's inner radius $\sim 10^7$ cm. One natural way for the BH to have the energy radiated at large radius $\gtrsim 10^9$ cm is that the BH and its accretion disk, which accrete at super-Eddington rates, eject a fraction of the supplied mass away. The radiation is advected with the outflowing, opaque gas until it can diffuse out at a larger radius. The idea of radiation-driven mass ejection has long been discussed for planetary nebulae (e.g., Faulkner 1970; Finzi \& Wolf 1971) and classical novae (e.g., Ruggles \& Bath 1979). This radiation-driven, optically thick (to continuum photons) outflow was considered for some ULXs and quasars with spectra dominated by soft, thermal component first by King \& Pounds (2003). King \& Pounds (2003) considered only the electron scattering opacity and the surface where the scattering optical depth $\tau_s= 1$. However, the $\tau_s=1$ surface is not the real region that physically determines the emission's spectral property, as the absorption opacity also plays a critical role. In addition, for estimating the mass flow rate of the outflow, King \& Pounds (2003) used the local escape speed at $\tau_s= 1$ as the flow's speed. However, as we will argue, the speed can be very different and both much lower and much higher speeds are possible. 

Here we revise the optically thick outflow model, adding the important implication of absorption and reconsidering the issue of photon diffusion. We then apply it to M101 ULX-1. Our goal is to consider the interplay of the electron scattering opacity and the absorption opacity in shaping the emergent radiation's spectrum. This allows us to provide analytical formulae to calculate more accurately the emission radius and the density at that radius from the known observables (luminosity and color temperature). It also enables us to put constraints on the outflow mass loss rate. The analysis presented here can be potentially applied to other ULX sources with thermally dominated spectra, and to other super-Eddington accreting sources such as stellar tidal disruptions and coalescences of compact objects. The relativistic and limb-darkening effects are discussed by Ogura \& Fukue (2013) and are found to be negligible except for relativistic outflows. We consider sub-relativistic outflows here.

The Letter is organized as follows. In \S \ref{sec:radii} we identify the physical processes that shape the radiative properties of an outflow and two associated characteristic radii. We review in \S \ref{sec:opacity} the radiative diffusion and relevant opacities. We determine in \S \ref{sec:wind} the emission radius and the gas density at that radius. We finish with a summary and a discussion of the outflow's speed and mass loss rate in \S \ref{sec:conclusion}. 

\section{The outflow model for M101 ULX-1}

We consider a hot, expanding mass outflow from an accreting BH. The outflow is optically thick to continuum photons, and the radiation pressure dominates over the gas pressure inside the flow, therefore, the outflow is radiation driven. The energy source is the accretion of material (inflow) near the horizon of BH. Since the radii that we will consider are all much larger than the BH event horizon, we approximate the outflow to be spherically symmetric. Deep inside the outflow, photons are trapped and advected with the gas. Only at a larger radius photons can diffuse out. In the physical parameter space of our interest, the opacity due to electron scattering is more important than that due to absorption (see later Eq. \ref{eq:rho-crit}), although both need to be considered for determining the spectrum of the emergent radiation.   

\subsection{Characteristic radii} 	\label{sec:radii}

For the two observables of a thermally dominated spectrum, the bolometric luminosity $L$ and the color temperature $T_{\rm col}$, we determine the radius from which the photons are emitted and the density at that radius. To do so it is useful to define two characteristic radii for the outflow. 

The first radius is the photon ``trapping'' radius $R_{\rm trap}$ at which the outflow's optical depth is $\tau(R_{\rm trap}) \equiv \int_{R_{\rm trap}}^{\infty} \rho \kappa dr = c/v$, 
where $\kappa$ is the total opacity due to both scattering and absorption and $v$ is the outflow velocity at $R_{\rm trap}$. Below $R_{\rm trap}$, the photon diffusion time is longer than the expansion time and photons are advected with the flow. Beyond $R_{\rm trap}$ photons start to diffuse out. This radius is also where the ratio of advective luminosity to diffusive luminosity is unity (the ratio is larger inside, and smaller outside).  

The second radius, the ``thermalization radius'' $R_{\rm th}$, is the photons' last absorption surface at which the effective absorption optical depth is unity (e.g., Rybicki \& Lightman 1979): 
\beq     \label{eq:tau*general}
\tau_{\nu}^*(R_{\rm th})= \int_{R_{\rm th}}^{\infty} \rho \sqrt{\kappa_{\nu}^a (\kappa_{\nu}^a+\kappa_s)} dr= 1,
\eeq
where $\kappa_{\nu}^a$ is the monochromatic absorption opacity and $\kappa_s$ is the electron scattering opacity. The last absorption surface $R_{\rm th}$ is frequency-dependent because of the frequency dependence of $\kappa_{\nu}^a$ (see Eq. \ref{eq:kappa-nua} below), i.e., lower-frequency photons have larger $R_{\rm th}$. Therefore, the emergent spectrum is a modified blackbody in which the part below the Wien peak flattens out from the Rayleigh-Jeans slope, while this effect hardly changes the Wien tail (Rybicki \& Lightman 1979; Shapiro \& Teukolski 1983). {The photons around the Wien peak dominate the radiation energy as well as the photon number. In addition, the observed color temperature is identified based on the Wien peak photons. Therefore we define} $R_{\rm th}$ to be the last absorption surface of those photons near the Wien peak, i.e., of energy $h\nu = kT_{\rm th}$.

Below $R_{\rm th}$, the photons of energy $h\nu = kT_{\rm th}$ are in local thermal equilibrium with the gas. Beyond $R_{\rm th}$, these photons may still scatter off electrons multiple times but they are not absorbed, thus, photons are neither created nor destroyed and they are not in thermal equilibrium with the gas. For the case of $R_{\rm trap} < R_{\rm th}$, the photons' energy does not change at $r>R_{\rm th}$; the color temperature $T_{\rm col}$ of the emergent radiation will be equal to $T_{\rm th} \equiv T(R_{\rm th})$, the local thermal equilibrium temperature at $R_{\rm th}$. 
For the other case $R_{\rm th} < R_{\rm trap}$, the color temperature will be set at $R_{\rm trap}$, because in this case adiabatic cooling continues to change the photons' energy.

Above $\max(R_{\rm th}, R_{\rm trap})$, inelastic scatterings could take place between the photons and the electrons (Comptonization). The electrons that have cooled adiabatically are out of thermal equilibrium with photons. However, the Comptonization only causes the photons to lose a negligible fraction of their energy because the energy density of radiation is much higher than that of electrons. The Comptonization will maintain the mean energy of electrons to be equal to that of photons, which is still set at $\max(R_{\rm th}, R_{\rm trap})$.
  
\subsection{Radiative diffusion and opacities} 	\label{sec:opacity}

We turn now to derive the relation between the two observables, $L$ and $T_{\rm col}$, and the two unknowns, the emission radius $\max(R_{\rm th}, R_{\rm trap})$ and the gas density at that radius.
As long as the radiation is in thermal equilibrium with the gas, i.e., up to $\max(R_{\rm th}, R_{\rm trap})$, the luminosity from the outflow is given by the radiative diffusion equation:
\beq    \label{eq:Lnu}
L_{\nu}= -4\pi R^2 \times \frac{4\pi}{3} \frac{\partial B_{\nu}(T)}{\partial \tau_{\nu}(R)},
\eeq
where $B_{\nu}(T)$ is the Planck function for the local temperature $T$, and $\tau_{\nu}(R)$ is the total optical depth of the atmosphere above $R$ due to both absorption and scattering,  
\beq      \label{eq:tau-nu}
 \tau_{\nu}(R)= \int_{R}^{\infty} \rho (\kappa_{\nu}^a + \kappa_s) d r.
\eeq

The scattering opacity is given by Thompson scattering: $\kappa_s= 0.2 (1+X)$ cm$^2$ g$^{-1}$, where $X$ is the hydrogen mass fraction. The absorption opacity includes contribution from free-free and bound-free processes, both having the same scaling on density and temperature: $\kappa_{\nu}^a \propto \rho T^{-7/2}$, and almost the same frequency dependence (the Kramer's Law): $\kappa_{\nu}^a \propto \nu^{-3}$ for $h\nu/kT \gtrsim 1$. One can write the total monochromatic absorption opacity as 
\beq    \label{eq:kappa-nua}
\kappa_{\nu}^a = C \rho T^{-7/2} f(u)~~ \mbox{cm}^2 \mbox{g}^{-1},
\eeq
where $C$ is a constant which also contains the dependence on the gas composition, {$\rho$ is in units of g cm$^{-3}$ and $T$ in K,} $u \equiv h\nu/kT$ is the normalized frequency, and the dimensionless function $f(u)$ accounts for the frequency dependence and is normalized as $f(1)=1$. Since we specify $R_{\rm th}$ to be the last absorption surface for photons of $u=1$, the detailed form of $f(u)$ is unimportant. 

The value of $\kappa_{\nu}^a$ at $u=1$ dictates the value of $C$. In order to have an accurate estimate of $C$, we use the tabulated monochromatic opacity data computed by the Opacity Project (Seaton et al. 1994). The publicly available data (Seaton 2005) is the sum of $\kappa_{\nu}^a$ and $\kappa_s$. We then subtracted $\kappa_s$ and also included the correction factor for stimulated emission. From the data for a series of $\rho$ and $T$ we confirmed the dependence of $\rho T^{-3.5}$. For the composition identical to that inferred for M101 ULX-1, $X=0$ and the mass fraction of metals $Z=0.008$ (Liu et al. 2013), the data gives $C \simeq 2.4 \times 10^{25}$.   

It should be noted that the bound-bound line absorption also contributes to $\kappa_{\nu}^a$. However, for conditions ($T$, $\rho$ and $Z$) that are relevant to M101 ULX-1, we find the lines are still sparse around $u=1$. Only at higher frequencies, e.g., $u \sim 10$, the lines are dense enough to blend, and therefore to substantially contribute to $\kappa_{\nu}^a$, but those frequencies are irrelevant to our analysis.

It is straightforward to see from Equations (\ref{eq:tau*general}, \ref{eq:Lnu} and \ref{eq:tau-nu}) that when the absorption opacity dominates over the scattering one [$\kappa_{\nu}^a(u=1) \gg \kappa_s$], $\tau_{\nu}(R_{\rm th}) \simeq \tau_{\nu}^*(R_{\rm th}) \simeq 1$, and Eq. (\ref{eq:Lnu}) gives $L_{\nu} \sim 4\pi^2 R_{\rm th}^2 B_{\nu}(T_{\rm th})$, i.e., the emergent emission is a blackbody. Therefore the emission radius takes the blackbody value $R_{\rm th} \simeq R_{\rm BB} \equiv [L/(4\pi\sigma T_{\rm col}^4)]^{1/2}$, where $\sigma$ is the Stefan-Boltzmann constant.

In the following we assume that the scattering dominates over the absorption ($\kappa_{\nu}^a \ll \kappa_s$). For the case of $R_{\rm trap} < R_{\rm th}$, $T_{\rm col}= T_{\rm th}$. Eqs. (\ref{eq:tau*general}), (\ref{eq:Lnu}) and (\ref{eq:tau-nu}) become
\beq    \label{eq:tau*}
\tau_{\nu}^*(R_{\rm th}) \simeq \int_{R_{\rm th}}^{\infty} \rho \sqrt{\kappa_{\nu}^a \kappa_s} dr = 1, 
\eeq
\beq     \label{eq:Ldiff}
L \simeq \frac{16}{3} \pi R_{\rm th}^2 \frac{\sigma T_{\rm th}^4}{\tau(R_{\rm th})},
\eeq
\beq     \label{eq:tau-s}
\tau(R_{\rm th}) \simeq \tau_s(R_{\rm th}) = \int_{R_{\rm th}}^{\infty} \rho \kappa_s d r.
\eeq
The bolometric luminosity $L$ in Eq. (\ref{eq:Ldiff}) is obtained by integrating Eq. (\ref{eq:Lnu}) over frequency at $R_{\rm th}$, and by approximating the differential by a finite difference  $\partial T^4/ \partial r \simeq - T^4/r$. It is evident from Eq. (\ref{eq:Ldiff}) that $R_{\rm th}$ must be $> R_{\rm BB}$ because $\tau(R_{\rm th}) > 1$. 

Since the density drops as $\rho \propto r^{-2}$ (for a coasting outflow whose $v$ is independent of $r$) or faster (for an accelerating outflow whose $v$ increases with $r$), the integrals in Eqs. (\ref{eq:tau*}) and (\ref{eq:tau-s}) can be carried out to give $\tau_{\nu}^*(R_{\rm th}) \simeq \rho(R_{\rm th}) R_{\rm th} \sqrt{\kappa_{\nu}^a \kappa_s} \simeq 1$ and $\tau_s(R_{\rm th}) \simeq \rho(R_{\rm th}) \kappa_s R_{\rm th}$.
Solving Eqs. (\ref{eq:kappa-nua} -- \ref{eq:tau-s}) gives
\beq     \label{eq:Rth-rho}
\mbox{for} ~~ R_{\rm trap} < R_{\rm th}: ~~ \begin{cases}
R_{\rm th} \simeq R_{\rm BB} \sqrt{\tau_s(R_{\rm th})}, &\\
\rho (R_{\rm th}) \simeq (\kappa_s R_{\rm BB})^{-1} \sqrt{\tau_s(R_{\rm th})}, &\\
\tau_s(R_{\rm th}) \simeq \left( \frac{3 T_{\rm th}^3 L \kappa_s^4}{16 \pi \sigma C^2} \right)^{1/5}.  
\end{cases}
\eeq

On the other hand if $R_{\rm th} < R_{\rm trap}$, $T_{\rm col}= T(R_{\rm trap})$. Replacing $R_{\rm th}$'s in Eqs. (\ref{eq:tau*} - \ref{eq:tau-s}) and approximating the integrals, we have
\beq      \label{eq:trap-tau*}
\tau_{\nu}^*(R_{\rm trap}) \simeq \rho R_{\rm trap} \sqrt{\kappa_{\nu}^a \kappa_s} < 1,\\ 
\eeq
\beq      \label{eq:trap-L}
L \simeq \frac{16}{3} \pi R_{\rm trap}^2 \frac{\sigma T_{\rm col}^4}{\tau(R_{\rm trap})},\\
\eeq
\beq      \label{eq:trap-tau}
\tau(R_{\rm trap}) \simeq \rho \kappa_s R_{\rm trap} = c/v.
\eeq
These can be solved to give $R_{\rm trap}$ and $\rho(R_{\rm trap})$ as functions of $v$ and a lower limit on $v$:
\beq     \label{eq:Rtrap-rho}
\mbox{for} ~ R_{\rm th} < R_{\rm trap}, ~~ \begin{cases}
R_{\rm trap} \simeq R_{\rm BB} \sqrt{c/v}, &\\
\rho (R_{\rm trap}) \simeq (\kappa_s R_{\rm BB})^{-1}  \sqrt{c/v}, &\\
v(R_{\rm trap}) > c \left( \frac{16\pi \sigma C^2}{3L T_{\rm col}^3 \kappa_s^4} \right)^{1/5} \equiv v_{\rm crit},
\end{cases}
\eeq 
where $v_{\rm crit}$ is the critical speed when $R_{\rm th}= R_{\rm trap}$. {The outflow mass rate in this case is $\dot{M} \equiv 4\pi R^2 \rho v = 4\pi c R_{\rm BB} \kappa_s^{-1} \sqrt{c/v}$.}

\begin{figure}
\includegraphics[width=8.5cm, angle=0]{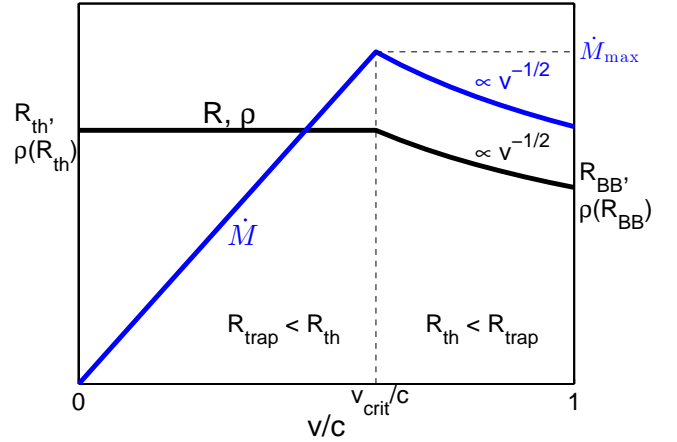}
\caption{Schematic plot of the solution of the emission radius and the density at that radius of an outflow, for given $L$ and $T_{\rm col}$ and under the condition that $\kappa_{\nu}^a \ll \kappa_s$. Also shown is the outflow mass rate.}    \label{fig1}
\end{figure}

Figure \ref{fig1} shows collectively the solution of the emission radius and the density as functions of $v$ under the condition $\kappa_{\nu}^a \ll \kappa_s$, for given $L$ and $T_{\rm col}$, for the both cases. It also shows the outflow mass rate. { $\dot{M}$ reaches its maximally allowed value $\dot{M}_{\rm max} = 4\pi c R_{\rm BB} \kappa_s^{-1} \sqrt{c/v_{\rm crit}}$ when $v= v_{\rm crit}$ (or $R_{\rm th}=R_{\rm trap}$).}

\subsection{Application to M101 ULX-1}		\label{sec:wind}

Now we apply the analysis for an outflow model to the high states of M101 ULX-1 to get the emission radius and the local mass density. The observables\footnote{Kong et al. (2004) reported a higher peak luminosity $\simeq 3\times10^{40}$ erg s$^{-1}$ for high states, but Mukai et al. (2005) argued that Kong et al. adopted an unphysically high absorbing neutral hydrogen column density; we follow here the value obtained by Mukai et al. (2005).} are $L \simeq 3\times10^{39}$ erg s$^{-1}$ and $T_{\rm col} \simeq$ 0.1 keV$/k$. {We consider first the case that scattering is unimportant ($\kappa_{\nu}^a \gg \kappa_s$) at $u= 1$. This requires} (cf. Eq. \ref{eq:kappa-nua})
\beq    \label{eq:rho-crit}
\rho(R_{\rm th}) \gg 1.4\times10^{-5} \mbox{g cm}^{-3}.
\eeq  
The emission radius is $R_{\rm th} \simeq R_{\rm BB} \simeq 1.3\times10^9$ cm, and $\tau_{\nu}^*(R_{\rm th}) \simeq \tau_{\nu}(R_{\rm th}) \simeq \rho \kappa_{\nu}^a R_{\rm th} \simeq 1$. The last relation gives $\rho(R_{\rm th}) \simeq 2.3\times10^{-7}$ g cm$^{-3}$, which is inconsistent with the condition of Eq. (\ref{eq:rho-crit}). Therefore, this case is ruled out for M101 ULX-1.

Turning now to the case that the scattering dominates over the absorption ($\kappa_{\nu}^a \ll \kappa_s$), if $R_{\rm trap} < R_{\rm th}$, we get from Eq. (\ref{eq:Rth-rho})
\beq
\begin{cases}
R_{\rm th} \simeq 6.8\times10^9 ~~ \mbox{cm}, &\\
\rho (R_{\rm th}) \simeq 2.0 \times 10^{-8} ~~ \mbox{g cm}^{-3}, & \\
\tau_s(R_{\rm th}) \simeq 27. &
\end{cases}
\eeq
{Their dependence on $L$ and $T_{\rm th}$ are $R_{\rm th} \propto L^{3/5} T_{\rm th}^{-17/10}$, $\rho(R_{\rm th}) \propto L^{-2/5} T_{\rm th}^{23/10}$ and $\tau_s(R_{\rm th}) \propto L^{1/5} T_{\rm th}^{3/5}$.}
The resultant $\rho$ satisfies the $\kappa_{\nu}^a \ll \kappa_s$ condition (the reversed relation in Eq. \ref{eq:rho-crit}). The outflow mass rate has an upper limit $\dot{M} < \dot{M}_{\rm max} \simeq 2.1\times10^{-4}$ $M_{\odot}$ yr$^{-1}$.

If $R_{\rm th} < R_{\rm trap}$, $v$ must be $> v_{\rm crit}$. This suggests upper limits on $R_{\rm trap}$ and $\rho(R_{\rm trap})$ according to Eq. (\ref{eq:Rtrap-rho}). On the other hand $v < c$ suggests lower limits ({we assume throughout the paper that the outflow is not relativistic}). Together we have 
\beq     \label{eq:appl-trap}
\begin{cases}
v(R_{\rm trap}) > 1.1\times10^4 ~~ \mbox{km s}^{-1}, &\\
1.3\times10^9 ~~ \mbox{cm} < R_{\rm trap} < 6.8\times10^9 ~~ \mbox{cm}, &\\
3.8\times10^{-9} ~\mbox{g cm}^{-3} < \rho(R_{\rm trap}) < 2.0\times10^{-8} ~\mbox{g cm}^{-3}. &
\end{cases}
\eeq
The range of $\rho(R_{\rm trap})$ lies much below the range in Eq. (\ref{eq:rho-crit}), thus, it satisfies the $\kappa_{\nu}^a \ll \kappa_s$ condition. The range of $v$ similarly suggests a range of the outflow mass rate: $3.8\times10^{-5}$ $M_{\odot}$ yr$^{-1} < \dot{M} < 2.1 \times10^{-4}$ $M_{\odot}$ yr$^{-1}$. The narrow ranges of these parameters constrain the conditions of the $R_{\rm th} < R_{\rm trap}$ case. The non-relativistic assumption $v<c$ also suggests an upper limit to the kinetic luminosity of the outflow $L_k \equiv \dot{M}v^2/2 < 1.1\times10^{42}$ erg s$^{-1}$.

\section{Discussion}    \label{sec:conclusion}

While ULXs are frequently suspected as IMBH candidates, the recurrent outbursting M101 ULX-1 was recently identified as a BH / W-R binary whose kinetic measurement suggests a stellar-mass BH. In this paper we discuss an optically thick outflow from a stellar-mass BH for M101 ULX-1. The parameter regime relevant to this source suggests that the electron scattering opacity dominates over the absorption opacity ($\kappa_{\nu}^a \ll \kappa_s$). For the case that the thermalization radius lies above the trapping radius ($R_{\rm trap} < R_{\rm th}$), we consider the interplay between the two opacities. This allows us to determine the outflow's emission radius ($\simeq 6.8\times10^9$ cm) and density ($\simeq 2.0 \times 10^{-8}$ g cm$^{-3}$), given the observed luminosity and color temperature of the thermal spectrum.

The case of $R_{\rm th} < R_{\rm trap}$ is not ruled out. {This case} requires a very high outflow speed $v(R_{\rm trap}) > 1.1\times10^4$ km s$^{-1}$. The emission radius and the density lie intermediate between the values determined otherwise in the $R_{\rm trap} < R_{\rm th}$ case and the values corresponding to when the emitting surface is a blackbody. {The outflow mass rate $\dot{M}= 4\pi R^2 \rho v$ is in a narrow range of $(0.4 - 2)\times10^{-4}~M_{\odot}$ yr$^{-1}$.}

One interesting result is that there is an upper limit of the outflow mass rate $\dot{M} \lesssim 2\times10^{-4}~~M_{\odot}$ yr$^{-1}$, which is general and independent of the location of $R_{\rm th}$ and $R_{\rm trap}$. To further determine $\dot{M}$ in the $R_{\rm trap} < R_{\rm th}$ case, we need $v(R_{\rm th})$. Unfortunately, no observational measurement of $v(R_{\rm th})$ is available. King \& Pounds (2003) used the local escape speed at the radius where $\tau_s= 1$ to estimate the outflow mass rate. If we follow this approach and use $v_{\rm esc}(R_{\rm th})= (2GM/R_{\rm th})^{1/2}$ $\simeq$ 6,300 $M_1^{1/2}$ km s$^{-1}$ where $M_1= M/(10 M_{\odot})$, it gives $\dot{M} \simeq 1\times10^{-4}~M_1 ~ M_{\odot}$ yr$^{-1}$. 

However, there is no obvious reason that $v(R_{\rm th})$ has to be comparable to or larger than $v_{\rm esc}(R_{\rm th})$. If the outflow is radiation driven, the equation of motion in the optically thick ($\tau_s >1$) region is 
\beq   \label{eq:motion}
{\partial \over \partial r} \left(\frac{v^2}{2}\right) + {1 \over \rho} {\partial P \over \partial r} + {GM \over r^2}= 0.
\eeq 
The pressure gradient term $(\partial P/\partial r)/\rho$ acts against the gravity and accelerates the flow. In the optically thin ($\tau_s < 1$) region, the radiative acceleration term $-\kappa_s L/(4\pi r^2 c)$ replaces the pressure gradient term in Eq. (\ref{eq:motion}), where $L$ is the diffusive luminosity. These two terms are equivalent in magnitude as it can be seen that in the region around $\tau_s= 1$ the luminosity is $L= -4\pi r^2 c/(\rho \kappa_s) \times (\partial P/\partial r)$. Therefore, as long as $L$ is Eddington or super-Eddington, the outflow may continue to accelerate well beyond $R_{\rm th}$.  

Nevertheless, in order to constrain $v(R_{\rm th})$ and, in addition, the outflow speed at infinity, a solution of the global dynamics of a radiation driven outflow that includes the effect of radiative diffusion is needed. Parker's adiabatic solar-wind solution (Parker 1965; also see Holzer \& Axford 1970) is not suitable as it does not include the radiative diffusion. {Such a solution is beyond the scope of this letter and will be discussed elsewhere.}

Two observational prospects of the massive outflow for ULXs exist. First, the velocity of the outflow can be directly measured by detecting absorption lines in X-rays, as the metals in the further part of the outflow (already optically thin to continuum) absorb photosphere photons by bound-bound transitions, as has been done for quasar PG1211+143 (Pounds et al. 2003). Two ULXs (NGC 5408 X-1 and NGC 6946 X-1) might have shown such line features, but the fit statistics are poor (Middleton et al. 2014). Second, the kinetic power of the outflow can be estimated by observing the emission signature from its eventual interaction, e.g., in shocks, with the environment such as a dense cloud. This is demonstrated recently by Soria et al. (2014) for one ULX in galaxy M83. 

\section*{Acknowledgments}

This work was supported by ERC grants GRBs and GRB/SN, by a grant from the Israel Space Agency and by the I-Core Center of Excellence in Astrophysics. R.-F. S. thanks Chris Matzner for discussion and comments on an early version of the paper. 



\end{document}